\documentclass{article}
\usepackage{titlesec}
\usepackage{algorithm2e}
\usepackage[utf8]{inputenc}
\usepackage{amsmath}
\usepackage{verbatim}
\usepackage{amsthm}
\usepackage[utf8]{inputenc}
\usepackage[T1]{fontenc}

\newtheorem{lemma}{Lemma}

\title{On Brooks' Theorem}

\author{Gopalan Sajith \thanks{E-mail: sajith@iitg.ac.in}\\
Dept. of Computer Science and Engineering,\\ 
Indian Institute of Technology,\\ Guwahati, INDIA-781039\\
\and
Sanjeev Saxena\thanks{E-mail: ssax@iitk.ac.in}\\
Dept. of Computer Science and Engineering,\\ 
Indian Institute of Technology,\\ Kanpur, INDIA-208 016}

\date{\today}

\begin{document}

\maketitle

In this paper, we give two proofs of Brooks' Theorem. The first is obtained by modifying an earlier proof, and the second 
by combining two earlier proofs. We believe these proofs are easier to teach in Computer Science courses.

~\\


\section{Introduction}

Brooks'  
theorem states that every graph in which the maximum degree of a vertex is $\Delta$ can be coloured with $\Delta$ colours, unless it is either a complete graph or an odd cycle.

Brooks' theorem has several proofs 
(see, e.g. \cite{CR}.). The most popular proofs are due to Lovasz \cite{L} and based on the Kempe chain argument \cite{MV}.

The proof of Melnikov and Vizing \cite{MV} and Wislon \cite{W} uses
contradiction. In this paper, we modify this proof. The modified proof
is constructive and implies a linear-time algorithm. This is described
in Section~2. Some of these techniques are also described in
\cite{SS}.

The proof of Lovasz \cite{L} assumes properties of block-cutpoint
trees and implies a linear-time algorithm. Bondy~\cite{B,BM} used a
result of Chartrand and Kronk~\cite{CK} in the second step of the
proof by Lovasz. Chartrand and Kronk~\cite{CK} show that every
connected nonseparable graph has a DFS tree, in which some node has
at least two children, unless the graph is 
complete, 
or a
complete bipartite graph or a circuit. Bondy's proof is again
non-constructive. Zajac \cite{Z} gave a new proof, which implies a
linear time algorithm. In this note, we combine the proofs of Zajac and
Bondy to get a proof that is almost as simple as Bondy's, and also
results in a linear-time algorithm. This proof is described in
Section~3. 

In the rest of this paper, we assume that $\Delta\geq 3$ (thus,
avoiding the case of cycle graphs). We now sketch the greedy method
for colouring \cite{BM}. If the graph has a vertex $v$ which is of
degree less than $\Delta$, then carry out DFS starting at $v$, which
becomes the root of the resultant DFS-tree. Pick the nodes of the
DFS-tree in post-order (children before parent) \cite{AHU}, and colour
each node with the minimum colour missing amongst its neighbours in
the graph. As the parent is coloured after the node, at each node
(except the root), at least one neighbour is not coloured, and hence at
most $\Delta-1$ colours are present in its neighbourhood; thus each
node, except the root, can be coloured. The root can also be coloured
as it has at most $\Delta-1$ neighbours (hence at most $\Delta-1$
colours in its neighbourhood). Thus, we need to consider only the case
where each vertex is of degree exactly $\Delta$.

\section{The First Proof}

In this section, we modify the proof of Melnikov and Vizing\cite{MV}
and Wilson\cite{W}. Some of these techniques are also described in
\cite{SS}.

Delete any vertex $v$. As the graph is no longer $\Delta$-regular, it
can be coloured with $\Delta$ colours in linear time. In the original
graph, all vertices except vertex $v$ are thus coloured. 
If some colour is absent at $v$, then $v$ can be 
coloured immediately with that colour.  
(We say that a colour $\sigma$ is absent at $v$, if none of $v$ 
or its neighbours is coloured $\sigma$.)

If no colour is absent at $v$, then as $v$ is not coloured, $v$ 
has a neighbour of each colour from $\{1,2,\ldots,\Delta\}$. Let us denote
the neighbour of colour $i$ by $v_i$.

Now, suppose a colour $\mu$ is absent at some $v_j$; that is, neither 
$v_j$ nor any of its neighbours is coloured $\mu$. In this case, we could 
recolour $v_j$  with $\mu$, thereby freeing up colour $j$ to be used on 
$v$. Hence, if such a recolouring is possible, we are done.  

Consequently, in the only remaining case of interest, each neighbour 
$v_i$ must be \emph{saturated}---that is, every colour other than its 
own appears in its neighbourhood. In particular, each $v_i$ has exactly 
one neighbour of every colour in 
$\{1,2,\ldots,\Delta\} \setminus \{i\}$.  

As the given graph $G$ does not contain a clique of size 
$\Delta+1$, there must exist two nonadjacent neighbours of $v$. Without 
loss of generality, let these be $v_1$ and $v_3$.
As vertex $v_i$ is of colour $i$, vertices $v_1,v_2,v_3$ are coloured $1,2$ and $3$ respectively. 

If $v_1$ and $v_3$ are in different $1-3$ components, by interchanging
colours $1$ and $3$ in one of those components, $v$ can be made to
have two neighbours of the same colour, and hence $v$ can be given
colour $1$ or $3$. Thus, we need to consider only the case when $v_1$
and $v_3$ are in the same $1-3$ component. If the $1-3$ component
containing $v_1$ and $v_3$ is not a simple path, then let $y$ be the
first vertex (from $v_1$) 
of degree greater than two in the
$1-3$ component. Then, as at least three neighbours of $y$ are
coloured the same ($1$ or $3$), at least one colour, say $\mu$, is
absent at $y$. Give colour $\mu$ to $y$ and interchange colours $1$
and $3$ in the $1-3$ path from $v_1$ to (but excluding) $y$. As $v_1$
is now coloured $3$, $v$ can be given colour $1$. Thus, we can assume
that the $1-3$ component containing $v_1$ and $v_3$ is a path.

Similarly, we may assume that $v_2$ and $v_3$ are in the same $2-3$
component, $v_1$ and $v_2$ are in the same $1-2$ component, and these
components are simple paths.

We next show that $v$ can be assigned a valid colour from $\{1,2,3\}$.
Let $P_{13}$ be the $1-3$ path between $v_1$ and $v_3$.
If not every vertex on this path has a neighbour coloured $2$, then
let $x$ be the first vertex (from $v_1$) 
with no neighbour of colour $2$.  We recolour $x$ with colour $2$, and
interchange colours $1$ and $3$ in the $v_1$-$x$ subpath of $P_{13}$.
Thus, colour $1$ becomes free at $v$ and can be used at $v$. Similarly,
process $P_{12}$, the $1-2$ path between $v_1$ and $v_2$, and 
$P_{23}$, the $2-3$ path between $v_2$ and $v_3$.  
Hence, we may assume that every vertex on each of the paths $P_{12}, P_{13},$ and $P_{23}$ has a neighbour of the third colour.

Next consider the case where edge $(v_1,v_3)$ is absent but edges
$(v_2,v_3)$ and $(v_1,v_2)$ are both present. As each $v_i$ has
exactly one neighbour of each of the $\Delta-1$ colours different from
$i$, $v_1$ and $v_3$ are the neighbours of $v_2$ that are coloured $1$
and $3$ respectively, and $v_2$ is the only neighbour of $v_1$ and
$v_3$ coloured $2$. Thus, we simultaneously recolour vertices $v_1$ and
$v_3$ with colour $2$ and vertex $v_2$ with colour $3$. As a result,
colour $1$ becomes free at $v$. So we can give colour 1 to $v$.

We are left with the case where edge $(v_1,v_3)$ is absent and at
least one of edges $(v_2,v_3)$ or $(v_1,v_2)$ is absent. Without loss
of generality, assume that edge $(v_2,v_3)$ is absent along with edge
$(v_1,v_3)$. (Edge $(v_1,v_2)$ may or may not be present). Then paths
$P_{13}$ and $P_{23}$ are nontrivial (in that they have intermediate
vertices). 

If any vertex on $P_{23}$ has two neighbours of colour $1$, then some
colour $\mu$ is absent at it. Recolouring that vertex with $\mu$
ensures that $v_2$ and $v_3$ are not in the same $2-3$ component.
Interchanging colours $2$ and $3$ in one of them solves the problem,
as we have seen before. 

So, assume that every vertex on $P_{23}$ has exactly one neighbour of
colour $1$.
 
Now interchange colours $1$ and $3$ in $P_{13}$.  As a result, $v_1$
gets coloured $3$ and $v_3$ gets coloured $1$, and they are still in
the same $1$-$3$ component. 
The condition that no colour is absent at each of $v_1$ and $v_3$  remains 
valid. Also, note that the neighbourhood of every vertex on $P_{23}$ remains intact.

Let $w$ be the neighbour of $v_3$ of colour $2$, and $P'_{23}$ be the
part of $P_{23}$ from $w$ to $v_2$. Let us interchange colours $2$ and
$3$ in $P'_{23}$. As a result, $v_2$ will get coloured $3$ (and $w$
too gets coloured $3$).

If edge $(v_1,v_2)$ is not present, then the new colouring is valid.
As both $v_1$ and $v_2$ are now coloured $3$, colour $2$ becomes free
and can be used at $v$.

If edge $(v_1,v_2)$ is present, then the above colouring is not valid.
Give colour 2 to $v_1$. In the previous colouring, $v_1$ was the only
neighbour of colour $1$ of $v_2$. So, $v_2$ now has no neighbour of
colour $1$, and thus can be given colour $1$. Colour $3$ remains free
and can be used at $v$. 

Since each edge on these three paths is examined at most twice, we
can colour $v$ in linear time.

\section{The Second Proof}

We combine elements from the proofs of Zajac \cite{Z} and Bondy
\cite{B,BM} to obtain a simpler proof.

Pick any vertex $v$ of $G$; as $G$ is not $K_\Delta$, $v$ has a pair
of nonadjacent neighbours $x$ and $y$. Run DFS starting at $x$, first
choosing edge $(x,v)$ and then edge $(v,y)$. 

Either the DFS tree is a simple path (Hamiltonian path) or the DFS
tree has a node with two children.

\noindent {\textbf{Case 1:}} (The DFS tree is a Hamiltonian path.) As 
$\Delta \geq 3$, $v$ must have a neighbour $z$ other than $x$ and $y$. 
As $x$ and $y$ are not adjacent, 
give
colour $1$ to both. 
As the DFS tree, by hypothesis, is a path, $z$ lies on this path.
Colour the vertices on the path 
starting from the child of $y$ to the vertex just before $z$ (leaving 
$z$ uncoloured for now) in that order.
When a vertex $w$ on this path is picked for colouring, its child would still be uncoloured, thereby ensuring that
a colour is absent at $w$, with which it can be coloured. Similarly, next
colour the vertices on 
the path starting from the last vertex (the only leaf) back to $z$ in that order.
When a vertex $w\not=z$ on this path is picked for colouring, its parent would still be uncoloured, thereby ensuring that
a colour is absent at $w$, with which it can be coloured.
As $v$ is a neighbour of $z$ and is uncoloured,
it is possible to colour $z$ with a valid colour too.
Finally, vertex $v$ can now be coloured,
because it has two neighbours of colour $1$.  

\noindent {\textbf{Case 2:}} (The DFS tree has a node with two
children.) Assume that $p$ is the first vertex with 
two children (say, $s$ and $t$).  

\begin{lemma}
\label{lem:separator}
If there is no edge from any proper ancestor of $s$ to any proper 
descendant of $s$ in the DFS tree, then $s$ is a separation point.
\end{lemma}

\begin{proof}
This is a direct consequence of standard DFS tree properties 
\cite{AHU}: if all edges incident on the descendants of $s$ are contained 
within the subtree rooted at $s$, then removing $s$ disconnects the 
graph into two components, one containing its ancestors and one 
containing its descendants.
\end{proof}

\noindent {\textbf{Case 2a:}} (Either $s$ or $t$ is a separation 
point \cite{B,L}; without loss of generality, let $s$ be one.)  

Remove $s$; 
the remnant graph has two components , $C_1$ and $C_2$. 
As the neighbours of $s$ in each component are of degree  $\Delta-1$, these
components can be coloured (see Introduction). As $s$ has at least one neighbour
in each component, the degree of $s$ in subgraph $C_i+s$ (for $i=1,2$) 
is at most $\Delta-1$. 
As $s$ has fewer than $\Delta$ neighbours 
in each subgraph, $s$ can be coloured in each using one of the $\Delta$ 
colours.
If $s$ is coloured $\alpha$ in $C_1$ and $\beta$ in $C_2$, then interchange
colours $\alpha$ and $\beta$ in $C_2$. As a result, $s$ is now coloured $\alpha$ in both
the components.
Kempe-component recolouring in $C_2$ can be carried out in $O(m)$ time by first identifying all vertices in $C_2$ that are coloured $\alpha$ or $\beta$, 
then interchanging their colours.
(Here, $m$ is the number of edges in the graph.)

\begin{lemma}
\label{lem:tworemoval}
If neither child, $s$ nor $t$ of $p$, is a separation point, then 
removing both $s$ and $t$ leaves the graph connected.
\end{lemma}

\begin{proof}
Since neither $s$ nor $t$ is a separation point, each of them has an 
edge from its subtree to some ancestor of $p$. Therefore, when $s$ and 
$t$ are removed, all remaining vertices are still connected to $p$ 
through these back edges.
\end{proof}

\noindent {\textbf{Case 2b:}} (Neither $s$ nor $t$ is a separation 
point.)  
By Lemma~\ref{lem:tworemoval}, removing $s$ and $t$ does not disconnect 
the graph. Moreover, there can be no edge between $s$ and $t$, since 
such an edge would be a cross-edge with respect to the DFS tree 
\cite{AHU}.  

Give colour $1$ to both $s$ and $t$. Now run DFS starting at vertex $p$ 
in $G-\{s,t\}$. Colour the vertices of the DFS tree in post order (using the adjacencies of $G$ so that no neighbour of $s$ or $t$ gets colour $1$). 
When a vertex is picked for colouring, its parent would yet be uncoloured,  
so at least one colour would 
always  be available to colour it.
As $p$ has two neighbours ($s$ and $t$) of colour $1$, a colour 
is available for $p$ when its turn comes.  

\medskip
\noindent
In both cases, Kempe-component searches and recolourings are invoked 
only when  synchronising  
colours across separated components. Each such 
operation requires at most a linear-time search using BFS/DFS 
to identify the components induced by the vertices of the two colours. Interchanging the two colours
will also take linear time.
So, the entire algorithm runs in $O(m)$ time.

\bibliographystyle{acm}

\end{document}